\RequirePackage{amsmath}
\RequirePackage{fix-cm}

\documentclass[journal]{IEEEtran}

\usepackage[pdftex]{graphicx}
\graphicspath{{pdf/}{jpeg/}}
\DeclareGraphicsExtensions{.pdf,.jpeg,.png}

\usepackage{comment}
\usepackage{textcomp}
\usepackage{makecell}
\usepackage{mathtools}
\usepackage{commath}
\usepackage{array,multirow}
\usepackage[normalem]{ulem}
\usepackage{amsmath,amssymb,amsfonts}
\usepackage{algorithmic}
\usepackage{graphicx}
\usepackage{textcomp}
\usepackage{xcolor}
\usepackage{verbatim}
\usepackage[numbers, sort]{natbib}
\usepackage{tabularx}

\newcolumntype{C}{>{\hsize=\dimexpr2\hsize+8.25\tabcolsep+\arrayrulewidth\centering\relax}X}
\newcolumntype{U}{>{\hsize=\dimexpr2\hsize+6\tabcolsep+\arrayrulewidth\centering\relax}X}
\newcolumntype{Q}{>{\hsize=\dimexpr1\hsize+1\tabcolsep+\arrayrulewidth\centering\relax}X}
\newcolumntype{S}{>{\hsize=\dimexpr1\hsize+1.1\tabcolsep+\arrayrulewidth\centering\relax}X}
\newcolumntype{O}{>{\hsize=\dimexpr1\hsize+18\tabcolsep+\arrayrulewidth\centering\relax}X}
\newcolumntype{P}{>{\hsize=\dimexpr1\hsize+9\tabcolsep+\arrayrulewidth\centering\relax}X}

\newcolumntype{Y}{>{\centering\arraybackslash}X}

\ifCLASSINFOpdf

\else

\fi

\begin{document}

\title{Deep Multitask Learning for Pervasive BMI Estimation and Identity Recognition in Smart Beds}


\author{Vandad Davoodnia, Monet Slinowsky, Ali Etemad
\thanks{V. Davoodnia, M. Slinowsky, and A. Etemad are with the Department of Electrical and Computer Engineering, Queen's University, Kingston, ON, K7L 3N6 Canada.}
}



\maketitle

\begin{abstract}
Smart devices in the Internet of Things (IoT) paradigm provide a variety of unobtrusive and pervasive means for continuous monitoring of bio-metrics and health information. Furthermore, automated personalization and authentication through such smart systems can enable better user experience and security. In this paper, simultaneous estimation and monitoring of body mass index (BMI) and user identity recognition through a unified machine learning framework using smart beds is explored. To this end, we utilize pressure data collected from textile-based sensor arrays integrated onto a mattress to estimate the BMI values of subjects and classify their identities in different positions by using a deep multitask neural network. First, we filter and extract ${14}$ features from the data and subsequently employ deep neural networks for BMI estimation and subject identification on two different public datasets. Finally, we demonstrate that our proposed solution outperforms prior works and several machine learning benchmarks by a considerable margin, while also estimating users' BMI in a ${10}$-fold cross-validation scheme.
\end{abstract}


\IEEEpeerreviewmaketitle

\section{Introduction} \label{sec:sec1}
There has recently been increased interest in the continuous monitoring of health information. Consumers now frequently purchase wearables and smartphone applications to monitor their physical activity, eating habits, and health factors like heart rate and weight. These bio-metrics offer an informative picture of a person's overall physical health. One such metric is the Body Mass Index (BMI). A significant change in BMI is not only a health risk factor, but a behavioral risk factor as well \cite{ford2001self}. While BMI has been shown to be strongly correlated to the quality of life, it is also a valuable tool used in clinical practices, for instance, in calculating drug dosages and public health care policy \cite{kent2017body}. BMI significantly affects the diseases for which a person is at risk, including myocardial infarction \cite{kennedy2005prognostic}, diabetes \cite{chiu2011deriving}, heart disease, and bone fractures \cite{gonnelli2014obesity}. As $39\%$ of the world's adult population were reported as overweight in $2016$ \cite*{who}, continuous weight and BMI monitoring is an indispensable tool to prevent obesity and track health interventions. 

Applications that provide feedback to users on exercise habits are now widely used, encouraging users to be more active through unobtrusive sensing. Unfortunately, similar devices and applications for \textit{automated}, \textit{ubiquitous}, and \textit{continuous} monitoring of BMI are not widely available. Wearables and smartphone applications for monitoring BMI require users to actively and constantly measure and log their information, making them only as ubiquitous as the amount of effort the user puts into tracking. An alternative is obtaining such metrics automatically through a bed sensor system connected to the internet of things (IoT), which can facilitate unobtrusive, continuous, and immediate measurements.

As BMI is an important metric in public health, the literature has extensively investigated methods of monitoring weight and BMI information. \cite{smith1997use} suggested that BMI should be routinely reported for obese adolescents, as opposed to weight alone. Later studies suggested that self-reported BMI is not sufficient due to observed underestimations in overweight populations \cite{flood2000use}, hence needing more reliable technical solutions for BMI monitoring \cite{schrader2010smartassist}. For instance, \cite{chatterjee2013persuasive} used a wireless sensor network in a home environment to capture daily living activity of older adults with diabetes in order to detect BMI.

In smart environments where certain metrics are monitored, a critical concept to consider is identity recognition, which will enable the estimated metrics to be associated with the right subject and communicated with them accordingly \cite{pentland2000face}. Moreover, in general, through recognition of identity in intelligent IoT-enabled environments such as smart homes, specific personalized settings can be set, based on which different elements of the environment can be tuned to create the optimum ambient conditions \cite{dey2000context, cook2007smart}.

\begin{figure}[t]
    \centering
    \includegraphics[width=0.98\columnwidth]{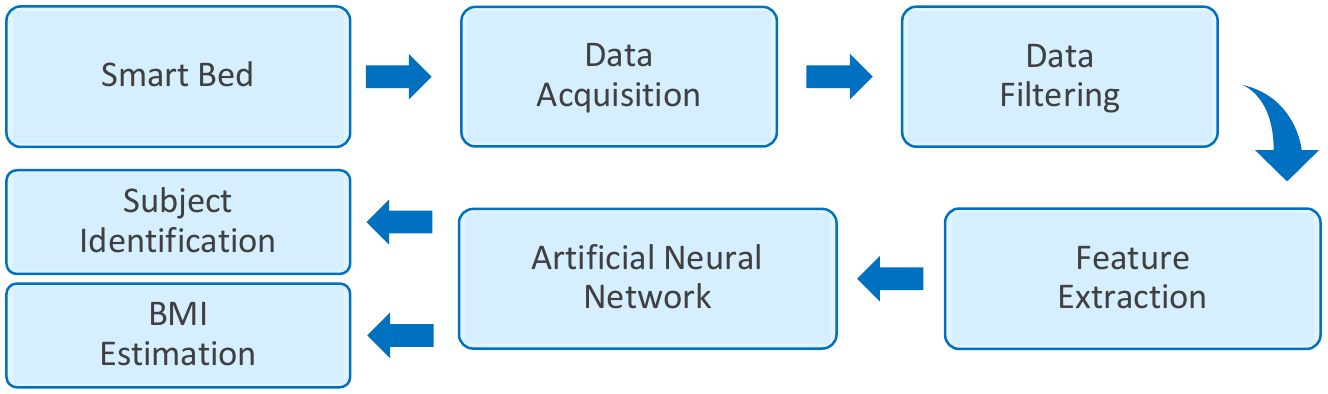}
    \caption{An overview of the work presented in this paper is illustrated. Data was first collected using a smart bed \cite{goldberger2000physiobank}, producing a set of pressure images. We then extracted features, which were fed into machine learning models to estimate BMI and identify subjects simultaneously.}
    \label{fig:overview}
\end{figure}

The development of pressure sensing textiles allows for sensors to be integrated into everyday objects, which can then record health information with minimal effort from the user. Pressure-sensing beds with built-in wireless capability are currently available on the market. These sensors can send data to the Cloud in real-time, where data can be securely stored and accessed by users from any device. Automated recognition and analysis of sleeping postures \cite{grimm2011automatic, davoodnia2019identity}, respiratory rate \cite{azimi2017breathing}, and identity \cite{pouyan2017pressure}, using such sensors have all been addressed in the literature. However, despite these advances, BMI values of users have not been estimated. Smart beds could eventually be an invaluable feature of smart homes. Such smart devices can also communicate with smartphones or smart watches, augmenting existing health informatics for a fuller picture of a user's health and physical condition. Additionally, this smart bed concept could be extended to hospitals. In a hospital setting, the information acquired through smart beds can be sent directly to nurses and doctors, as well as patient information databases, allowing for automated recording of the desired information.

In this paper, we propose and implement a method for simultaneous estimation of BMI and classification of identity in smart beds. We hypothesize that BMI values and features that allow for it to be estimated from pressure distributions are fairly unique to different individuals, especially in small populations often found in smart homes. As a result, through simultaneous learning of the feature manifolds for both BMI and identity, the model will be enforced, allowing for both tasks to be performed with high accuracy. Accordingly, we propose the use of a deep multitask neural network successive to noise reduction and features extraction from the original pressure data. We show that our method outperforms the state-of-the-art models reported in previous studies. An overview of our work is illustrated in Figure \ref{fig:overview}. The applications of our work include smart homes and clinics, where real-time and unobtrusive recording and tracking of BMI can be achieved while identification may be used for personalizing ambient smart home elements. Our contributions are summarized as follows:
\begin{itemize}
    \item We propose an accurate, automated, and ubiquitous method for estimating BMI in smart beds, for the first time;
    \item We propose a multitask neural network for simultaneous subject identification and BMI estimation, exploiting the internal relationship between the users and their BMI from the learned pressure representations;
    \item We evaluate our model on two public datasets, setting new state-of-the-art values for pressure-based subject identification and BMI estimation.
\end{itemize}

\section{Related Work} \label{sec:sec2}
Analyzing data obtained from pressure-sensing textiles is a relatively new area of research. Nonetheless, a variety of different characteristics and attributes such as posture, heart rate, and even age can be extracted regarding subjects when lying \cite{azimi2017breathing}, sitting \cite{damousis2008unobtrusive}, and standing \cite{javaid2017balance} on such pressure sensors. Here we discuss previous research utilizing pressure-sensing arrays in different contexts and towards different applications.

\subsection{Smart chairs} \label{sec:sub1sec2}
Posture detection has been carried out using chairs outfitted with pressure sensors \cite{meyer2010design} and force and acceleration sensors \cite{zemp2016application}. These studies classify a user's data into a number of known posture classes. For instance, \cite{meyer2010design} trained a Na\"ive Bayes classifier (NB) on a number of features extracted from a seat pressure array to distinguish such postures. The nature of the posture and a user's changes in posture have then been used to identify affective states as well. One study determined workers' stress levels based on how they sit in their office chair using self-organizing maps (SOM) and XY-fused Kohonen Networks on a set of extracted features from pressure maps \cite{arnrich2010does}. Another study utilized pressure information to determine students' interest levels in a learning environment by a set of independent Hidden Markov Models (HMM) \cite{mota2003automated}. Moreover, a study used a sensing seat for authenticating users' identities, with applications for identifying imposters in airports, laboratories, and other high-security buildings \cite{ferro2009sensing}. These sensing seats require users to assume a particular posture for a short period of time to verify their identity. Pressure sensors have also been incorporated into larger systems along with medical, audio, and visual sensors. A chair capable of pervasively sensing heart rate using an electrocardiogram (ECG) on the armrest and respiratory rate using pressure sensors has been developed \cite{griffiths2014health}. However, the study identified numerous flaws. First, the ECG used to sense heart rate required skin-to-skin contact, which was obtrusive to users. Second, few effective heart rate readings could be performed with the armrests as users used them sporadically-if at all.

\subsection{Smart bed applications} \label{sec:sub2sec2}
Though chair-based pervasive sensing has merit in some applications, for others it may be inconvenient. Depending on a user's occupation, they may sit in a chair sporadically in different positions and postures depending on the task. Conversely, a person lays in bed for $6$-$12$ hours a day, often in the same position, with little movement. Therefore, beds offer an ideal sensing environment. A number of metrics have been extracted and analyzed using sensor-enabled beds in the past. In a bed environment, posture detection has been given a great deal of attention as it is a metric in assessing sleep quality \cite{grimm2011automatic} and sleep quality is a valued indicator of overall health. Posture information has also been used to identify bedridden patient movement intentions, which an automated bed can fulfill \cite{chica2012real}. \cite{chica2012real} designed and developed an automated bed, embedded with pressure sensors, and performed extensive experiments by collecting data from $20$ subjects. Finally, they compared their proposed Multi-Layer Perceptron (MLP) model with $5$ other classifiers and reported a movement intention accuracy of $87\%$. Another study showed that sensor-enabled beds also show promise in monitoring and preventing ulcers in bedridden patients in a hospital \cite{yousefi2011smart}. This work reported a posture detection accuracy of $97.7\%$ using a Support Vector Machine (SVM) based model to distinguish between $5$ sleeping postures, and finally performed a risk assessment study on their data. Similarly a more recent study using pressure mattresses reported an average posture recognition accuracy of $75\%$ using a $4$ layer MLP \cite{enayati2018sleep}. In a more recent study, \cite{davoodnia2019bed} performed pose estimation by training a deep pre-processing network using pressure data captured by \cite{pouyan2017pressure}, achieving a limb detection rate of $95\%$.

Pressure images clearly show which parts of the body are under the most pressure in any given position, allowing health care workers to confidently position patients such that no part of the body is under pressure for any significant length of time. Similarly, pressure sensing has been used to classify infant state (crying, moving, or quiet), posture, and movement \cite{harada2000infant}, which has applications in decreasing risks of Sudden Infant Death Syndrome (SIDS). Beyond posture, bed sensors have been used to detect biological signals in real time. The respiratory rate has been monitored using bed pressure arrays \cite{jones2006reliable} and combinations of pressure sensors and microphones \cite{nishida2002non}. Smart beds have been shown to be an effective, non-invasive means to measure respiratory rate, outperforming some common methods \cite{azimi2017breathing}. Preliminary work has also been done in using smart beds to detect heart rate by observing the temporal power spectrum of pressure sensors \cite{spillmanjr2004smart}.

\subsection{Subject Identification} \label{sec:sub3sec2}
Subject authentication and identification using pressure sensors have recently been receiving attention. For instance, \cite{zhang2018ballistocardiogram} recorded Ballistocardiogram (BCG) and Electrocardiogram (ECG) of $91$ subjects while sitting on a chair. They use a long-short-term-memory (LSTM) network to achieve an average subject identification accuracy of $97\%$, $98\%$, and $100\%$ while using BCG, ECG, and the union of both signals respectively in an intra-subject validation scheme. Similarly, \cite{heydarzadeh2017gaits} studied gait pressure sensors to identify walking subjects using an SVM with a radial basis function (RBF) kernel. They reported an accuracy of $97\%$, outperforming prior works by $8\%$.

Estimation of BMI with pressure-sensing smart beds has not been addressed in the literature to the best of our knowledge. Additionally, identification of users in such settings has only been tackled by \cite{pouyan2017pressure}, who collected pressure distribution data from $13$ participants in $17$ postures. They extracted statistical features from the pressure distribution images and used them to train three separate deep learning models in three postures for identity recognition. It appears that the various postures presented a barrier to user identification as the work considered only the three coarse posture categories, namely right, left, and supine, with a separate neural network for each. In this paper, we propose a novel multitask learning solution to identify users regardless of posture using the dataset published in \cite{pouyan2017pressure}.

\section{Materials and Methods} \label{sec:sec4}
\subsection{Datasets}
\begin{figure}
    \centering
    \includegraphics[width=0.95\linewidth]{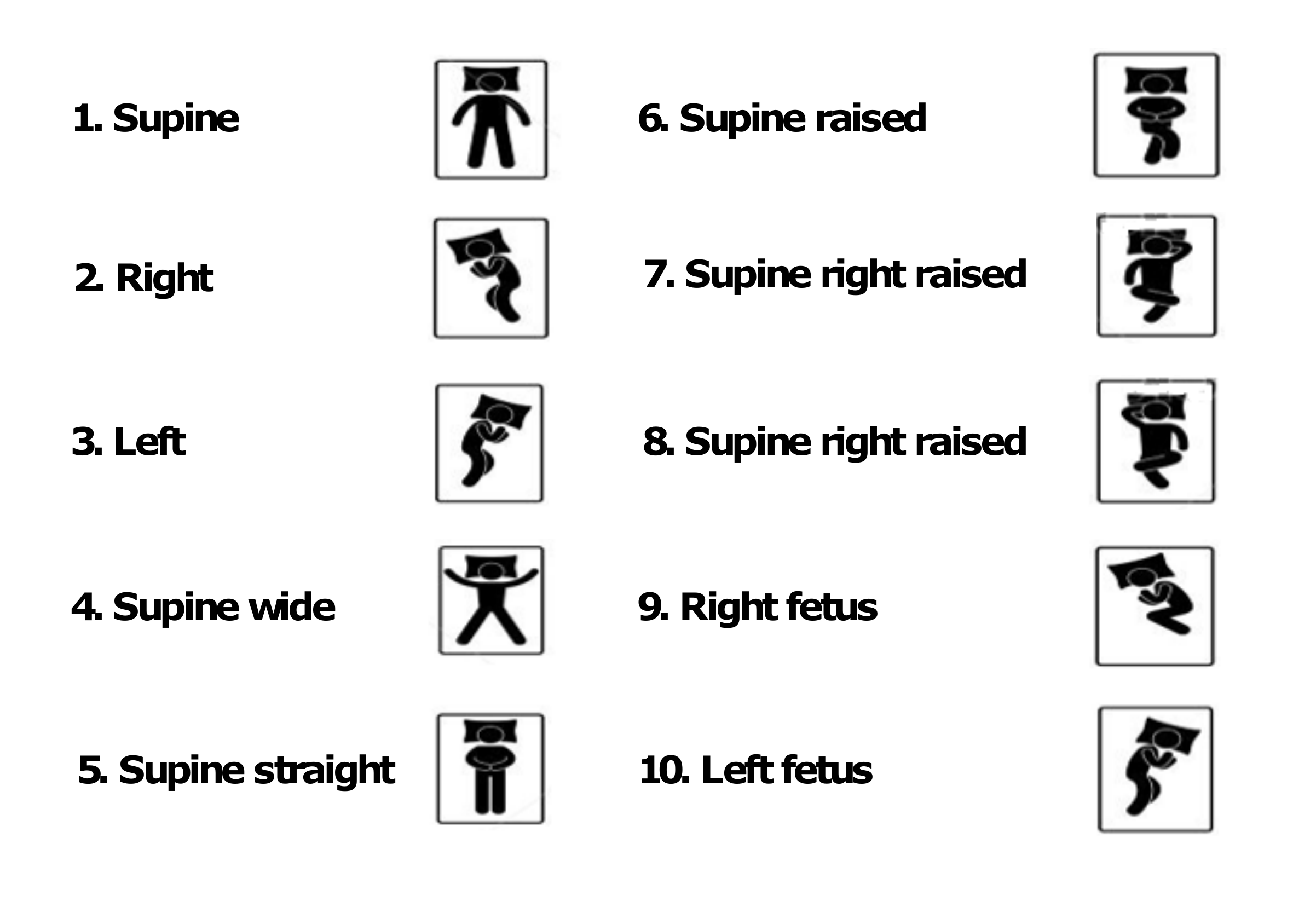}
    \caption{The $10$ common sleeping posture groups used for the study are presented. Participants assumed the postures while lying flat on a pressure sensing mattress. Posture images were reproduced from \cite{pouyan2017pressure}.}
    \label{fig:2}
\end{figure}

The data for this work were obtained from two public datasets, namely PmatData \cite{pouyan2017pressure} available in the PhysioNet repository \cite{goldberger2000physiobank} and HRL-ROS dataset \cite{clever20183d}. The details of these datasets are as follows:

\textbf{PmatData} \cite{pouyan2017pressure} contains pressure data from $13$ subjects, measured using a pressure mapping mattress. The mattress covered an entire queen size bed and was thin, light, and flexible. Each sensing mat consisted of a $32 \times 64$ array of pressure sensors with a scan rate of $3072$ sensors/second. The sensor points were uniformly distributed $1$ inch apart on the mattress, with a sampling frequency of $1.5$ Hz. Thirteen healthy individuals with no history of sleeping disorders participated in the study. The subjects assumed $17$ different sleeping postures for about two minutes each. The height, weight, and age of each participant were also provided. In seven of the positions, the participants lied on sleeping wedges at different angles. We chose to use the ten non-wedged positions and include the wedged postures in the same categories, as these most closely reflect the reality of the smart bed environment. Consequently, we used a total of $18256$ pressure map samples in our study. The ten non-wedged postures are shown in Figure \ref{fig:2}, showing that each position inherently places different amounts of pressure on the bed. These $10$ postures reflect the most frequently assumed natural poses.

\textbf{HRL-ROS} \cite{clever20183d} has been collected for $3$D human pose estimation with smart beds. The authors collected data using a motion capture system and a configurable bed equipped with pressure sensing arrays. During data collection, the subjects were asked to sit or lie in $6$ and $13$ different postures respectively, and move a limb in a specific path, while motion capture cameras were simultaneously used to record joint locations. To collect the pressure maps, the bed utilized a $27\times64$ grid of pressure sensors, distributed $2.86$\textit{cm} apart. The HRL-ROS dataset consists of $17$ subjects with a height range of $160-185$\textit{cm}, a weight range of $45.8-94.3$\textit{Kg}, and an age range of $19-32$ years. For this research, we used all the \textit{lying} data, adding up to a total of $39095$ pressure maps.

\subsection{Pre-Processing}
\begin{figure}[t]
    \centering
    \includegraphics[width=0.95\linewidth]{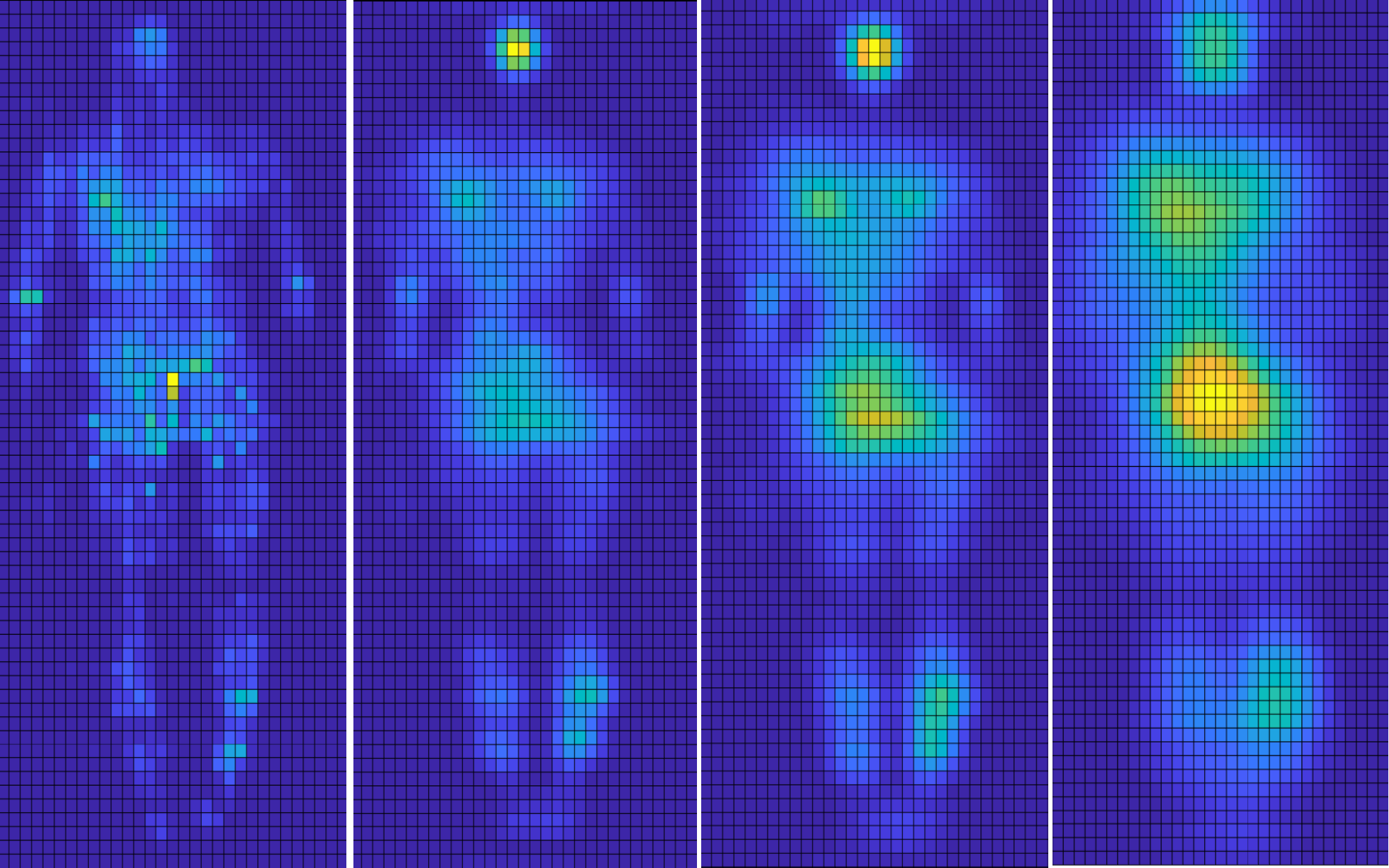}
    \caption{The effect of window size when applying the spatial filter is illustrated. The first image on the left presents the input unfiltered frame, while the next three present the impact of $2 \times 2$, $3 \times 3$, and $6 \times 6$ filter window sizes respectively.}
    \label{fig:windowsizes}
\end{figure}

In the pre-processing step, we sought to eliminate the noise caused by participants' voluntary and involuntary movements such as motion artifact, as well as the noise produced by other factors such as static buildup and malfunctioning sensors. Sample pressure images from the PmatData before and after filtering are shown in Figure \ref{fig:windowsizes}. In this figure, each pixel corresponds to a pressure sensor and lighter colors indicate higher pressure values. To reduce the noise without diluting the pressure information, we first applied a spatial median filter to each frame. We experimented with several window sizes from $1 \times 1$ to $20 \times 20$ and found that the $3 \times 3$ size produced the optimal outcome. Window sizes smaller than this had little effect on the image, whereas larger sizes blurred the image, making it difficult to visually distinguish between body shapes or postures as seen in Figure \ref{fig:windowsizes}. Next, we filtered the data along the time axis to reduce fluctuations in sensor values over time. We used a Gaussian filter for each sensor to generate a smooth pressure image suitable for the feature extraction step. The Gaussian window size was set to $5$. Similar to the spatial filter, the size of the Gaussian filter was selected through visual inspection and experimentation with the goal of maximizing performance. Our experiments showed that in the absence of these filters, the BMI estimation and identity classification performances presented in the following sections drops.

\subsection{Feature Extraction}
\begin{figure}[t]
    \centering
    \includegraphics[width=0.93\linewidth, scale=0.95]{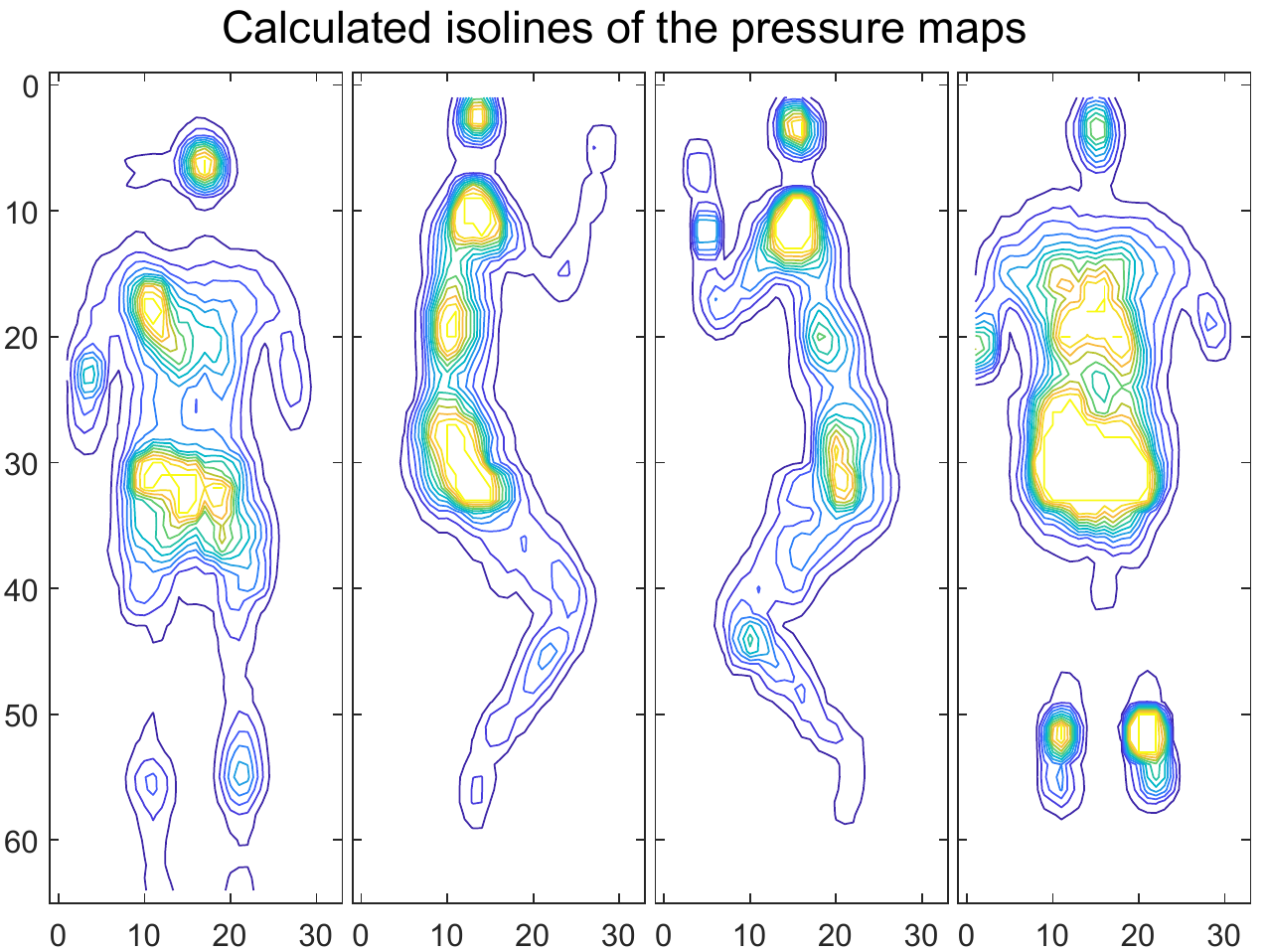}
    \caption{Examples of the calculated isolines of the pressure maps from PmatData is presented in this figure, where the definition of the lines provides two additional features.}
    \label{fig:3}
\end{figure}

We extracted $14$ different features for each frame. Statistical features consisting of maximum, mode, entropy, and range were measured along with mean, variance, skewness, and kurtosis. The equations for these features are presented in Table \ref{eq_table}, where $s_i$ is the value of sensor $i$ at a given time, $\mu$ is the mean, $\sigma$ is the standard deviation, $g$ is the skewness, $N$ denotes the non-zero pressure points in a frame, $k$ denotes the kurtosis, $e$ is entropy calculated using the estimated probability $P_i$ for 256 bins, and $E$ is the expected value. As shown in the table, the number of sensors with non-zero values in each frame was also extracted as a feature. Three additional features were extracted by counting the number of sensors with values between specific thresholds, where $\tau_1 = 20$, $\tau_2 = 60$, and $\tau_3 = 100$. These values of $\tau$ empirically showed the best performance. The final two features were properties of the contour matrix of each frame. First, we selected a maximum of $20$ contour levels divisible by $2$ or $5$ in the range $[\min(S), \max(S)]$, where $S$ is the matrix of pressure sensors. This resulted in a list of contour levels $C = {c_1, c_2,...c_n}$. Starting with $c_1$, the minimum contour level, a marching squares algorithm was used to find the isolines. The algorithm divides the $S$ matrix into squares of $2 \times 2$ pixels and traverses the square edges. If $c_i$ is in the square, linear interpolation is performed:
\begin{equation} \label{equ:feq10} 
\ r = \frac{c - p_1}{p_2 - p_1} ,
\end{equation}
\begin{equation} 
\label{equ:feq11}  
\ x_c = x_1 + r\times(x_2 - x_1) ,
\end{equation}
where $r$ is the slope of the contour line, $x_c$ and $y_c$ define the isoline at level $c$, $p_1$ and $p_2$ are the edges of the square, and $(x_1, y_1)$ and $(x_2, y_2)$ denote the location of $p_1$ and $p_2$ in the sensor matrix. The remaining squares are traversed until the line returns to its initial point or reaches the end of the grid. Then, the procedure repeats for the next contour level. The number of isolines in a frame varies depending on the range of sensor values. Effectively, each line shows an area of constant pressure. This operation segments the images as shown in Figure \ref{fig:3}. We used the size of the contour matrix (i.e. the number of isolines) as one feature and the sum of the $x_c$ and $y_c$ values as another feature (see Table \ref{eq_table}).

\newcommand\T{\rule{0pt}{4ex}}       
\begin{table}[t]
\caption{List of extracted features and their equations.} \label{eq_table}
\centering
\begin{tabularx}{\columnwidth}{l|Y} 
\hline
Name & Formula \\ \hline    
Max                &    $ \label{equ:feq12}  Max(s)                                  $  \T  \\ 
Mode               &    $ \label{equ:feq13}  Mode(s)                                $   \T \\
Range              &    $ \label{equ:feq14}  Range(s)                                $  \T  \\     
Entropy            &    $ \label{equ:feq17}  e=-\sum_{i}^{256}{P_i\log{P_i}}                                 $  \T  \\                  
Mean               &    $ \label{equ:feq1}  \ \mu = \frac{\sum_{i=1}^{N} s_i}{N}                      $  \T  \\
Variance           &    $ \label{equ:feq2}  \sigma^2 = \frac{\sum_{i=1}^{N}{(s_i-\mu)^2}}{N}          $  \T  \\
Skewness           &    $ \label{equ:feq3}  \ g = \frac{\sum_{i=1}^{N} (N_i - \mu)^3 /N)}{\sigma_i^3} $  \T  \\ 
Kurtosis           &    $ \label{equ:feq4}  \ k= \frac{E(s - \mu)^4}{\sigma^4}                        $  \T  \\
Non-Zero Count     &    $ \label{equ:feq5}  N                                                         $   \T  \\
20-60 Count        &    $ \label{equ:feq7}  N_{20<s<60}                                 $  \T  \\
60-100 Count       &    $ \label{equ:feq8}  N_{60<s<100}                               $ \T   \\
Above 100 Count    &    $ \label{equ:feq9}  N_{100<s}                                 $  \T  \\
Number of Isolines &    $ \label{equ:feq15} C                                          $ \T  \\
Sum of Isolines    &    $ \label{equ:feq16} \sum_{c}^{C} x_c+y_c                     $  \T  \\ \hline  
\end{tabularx}
\end{table}

\subsection{Deep Multitask Neural Network}
\begin{figure}[t]
    \centering
    \includegraphics[width=0.95\linewidth]{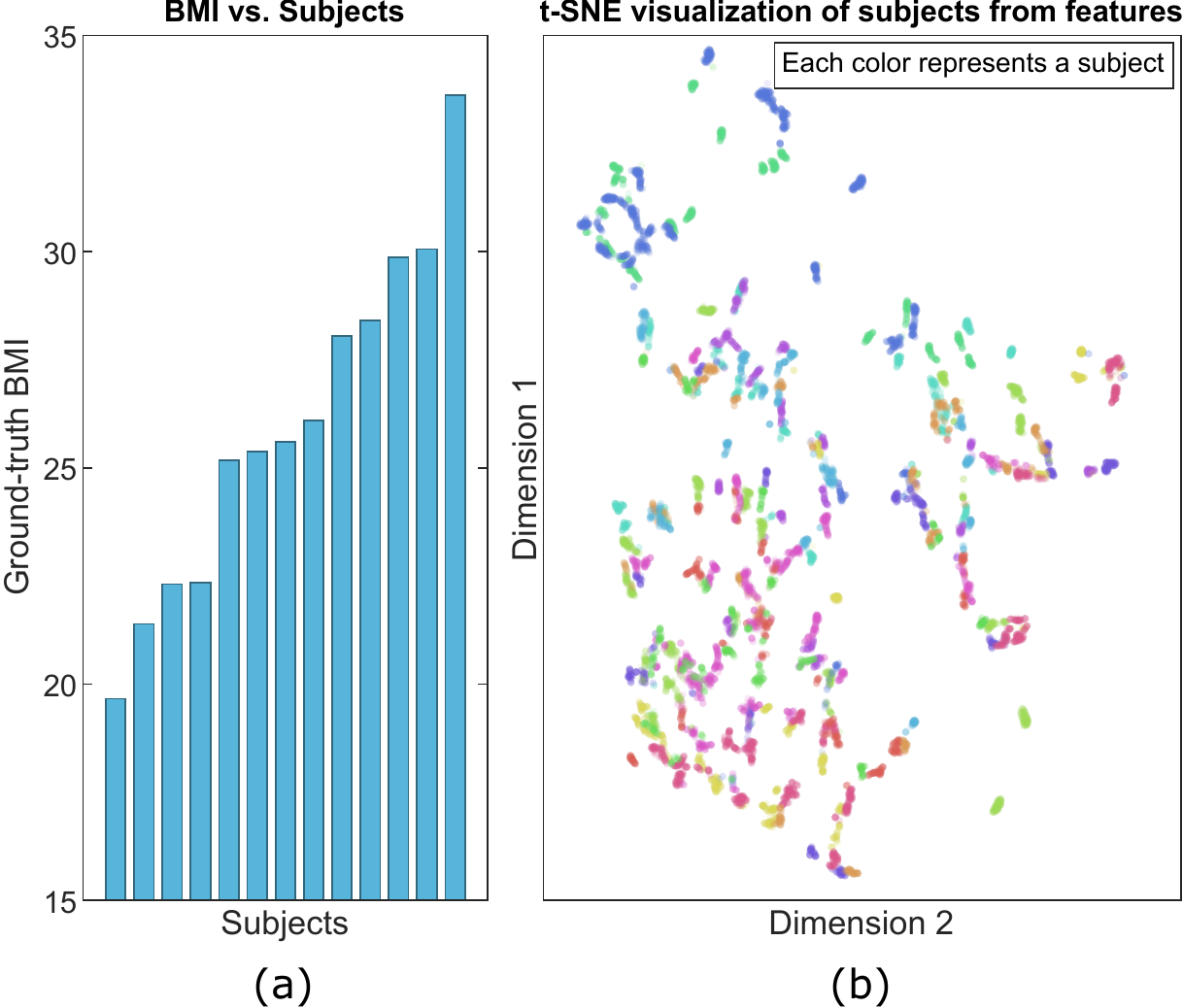}
    \caption{(a) The participants' BMI values for PmatData are illustrated, showing relatively unique measurements for each subject. (b) The t-SNE visualization of subjects over our feature space is illustrated, showing that each subject forms fairly separated clusters, indicating the potential for subject identification using the original feature space.}
    \label{fig:tsne}
\end{figure}

The utilized datasets contained both the height and weight of the subjects, which we used to calculate the ground-truth BMI values of the subjects. We hypothesize that pressure distributions used to estimate BMI in a learned model are fairly unique to each subject, especially in smart environments in which the number of subjects is constrained to a limited number (e.g. smart homes). Figure \ref{fig:tsne} (a) shows that, indeed, in the PmatData dataset, different subjects have different BMI values, and therefore, it is highly likely that the underlying features used to learn BMI and identity can reinforce one another. Moreover, in Figure \ref{fig:tsne} (b), we utilized t-Distributed Stochastic Neighbor Embedding (t-SNE) to visualize the features described earlier, projected onto a 2D space. It can be seen that the different subjects are fairly distinct in the new reduced space. As a result, we propose a multitask learning approach for estimating BMI and performing identity recognition simultaneously.

\begin{figure}[t]
    \centering
    \includegraphics[width=0.9\columnwidth]{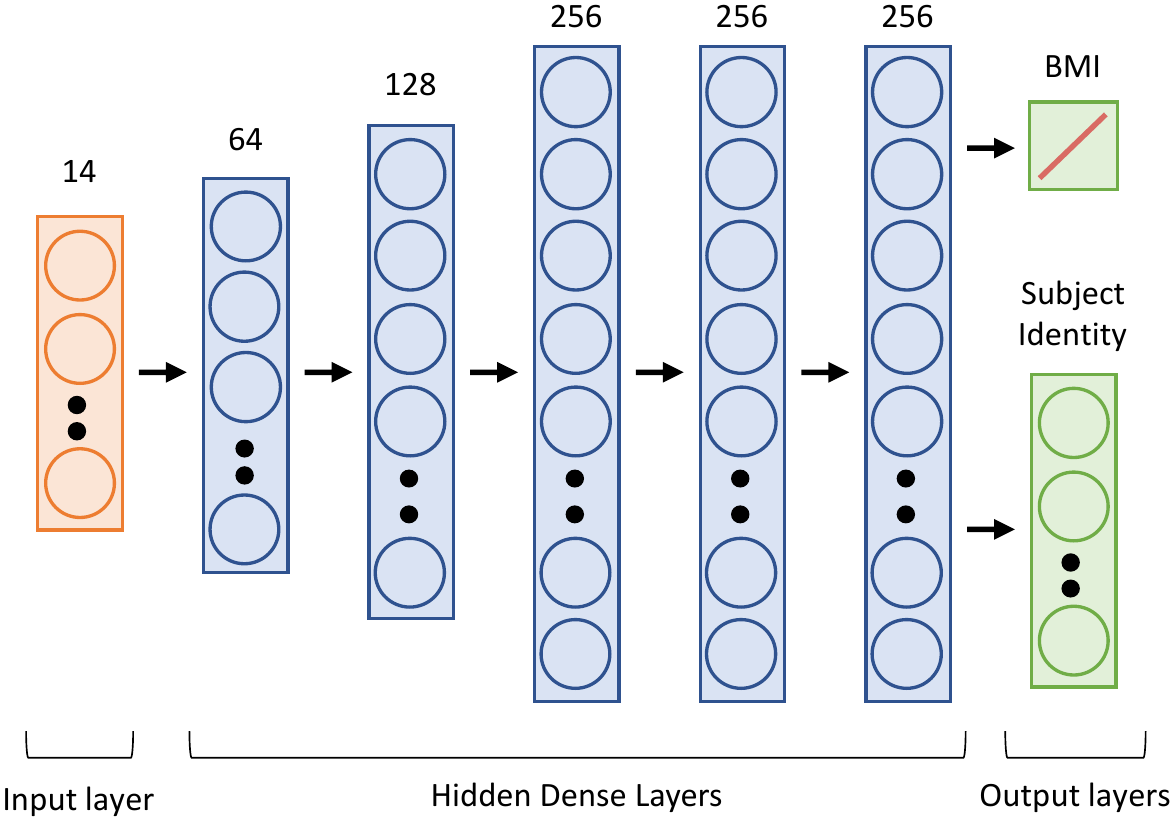}
    \caption{The neural network architecture is illustrated. The network consists of $14$ input neurons, five hidden dense layers, and $2$ output layers for BMI and subject identification.}
    \label{fig:neuralnet}
\end{figure}

Our proposed method uses a deep multitask neural network as a generative model for BMI estimation and a discriminative model for subject identification. Specifically, we utilized an MLP, which is highly effective for estimation purposes. MLPs are structured as feed-forward networks with one input layer, a number of hidden layers, and an output layer. The network architecture can be seen in Figure \ref{fig:neuralnet}, where $14$ input neurons and five hidden dense layers, containing $64$, $128$, $256$, $256$, and $256$ neurons with $tanh$ activation function. The final estimation/classification is then achieved by feeding the fifth hidden layer to two dense layers placed in parallel in the output layer, first one for BMI estimation using a linear activation, and the second one using a SoftMax activation to calculate the subject's class probability.

Let $S \in \mathbb{R}^{\textit{F}}$, with $F=14$, be a dataset made of features from pressure sensor maps containing a set of subjects identities $I_S$ with a specific BMI, $B_S$. The objective is to learn a function $\Phi(S; \theta) : \mathbb{R}^{\textit{F} \times \abs{\theta}} \rightarrow \mathbb{R}^\textit{M+N}$, where $\theta$ is the set of trainable parameters, $M$ and $N$ are the number of subjects and BMI target values in the dataset, with $M = \abs{I_S}$ and $N=1$ respectively, and $\Phi$ transfers the feature space $S$ onto a representational space, wherein one, the BMI weights are estimated and in another, the subjects identity probabilities are predicted. To address this problem, we considered training a deep neural network (DNN) to discriminate between subjects and estimate the BMI in $S$. To train this network, we utilize a set composed of tuples $(S, i, b)$ where $S$ represents our features and $i$ and $b$ are the subject identity and their corresponding BMI values respectively, that is, $i \in I_S$ and $b \in B_S$. Accordingly, we created a DNN, with the objective in mind. Therefore, The last layer of the neural network consists of a single multinomial logistic regressor for subject identification and one other node for BMI estimation, placed in parallel, with $M$ and $N$ units respectively. The output layers then estimate $P(i|S)$ and $BMI(b|S)$ accordingly. To learn both functions, we consider using the multi-class cross-entropy loss of the subjects and mean square error of BMI for data sample $S$ as follows:
\begin{equation}
\begin{aligned}
\mathcal{L}_{subject} & = -\sum_{j=1}^{M}{i_{j} \log P(i_j | S)}, \label{equ:equ1}\\ 
\end{aligned}
\end{equation}

\begin{equation}
\begin{aligned}
\mathcal{L}_{bmi} & = \frac{1}{2}{(b_{j} - BMI(b_j | S))}^2. \label{equ:equ2}\\ 
\end{aligned}
\end{equation}

For training our model, we consider a summation of both losses to learn multi-tasking as follows:
\begin{equation}
\begin{aligned}
\mathcal{L}_{total}& = \mathcal{L}_{subject}+\mathcal{L}_{bmi}, \label{equ:equ3}
\end{aligned}
\end{equation}
where $\mathcal{L}_{total}$ is then minimized in order to learn the weights according to our objective. We combined both losses as is, since applying a weighted sum of them did not make a significant improvement to the performance of either task.

\begin{figure*}[t]
    \centering
    \includegraphics[width=1\textwidth]{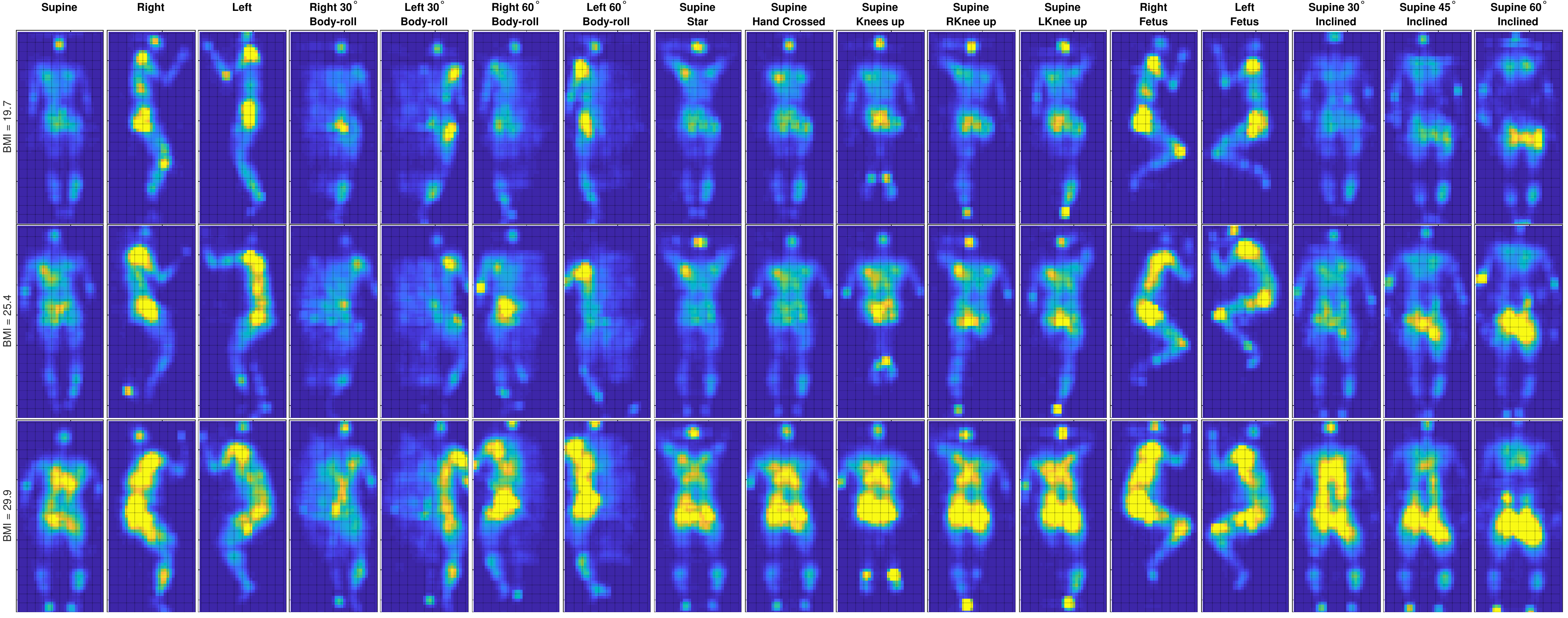}
    \caption{The pressure images of $3$ subjects in all the $17$ postures from PmatData is illustrated. Row $1$, $2$, and $3$ present the pressure matrices for a subject with a BMI of $19.7$, $25.4$, and $29.9$ respectively. Columns depict the different postures. Blue pixels indicate sensors with low-pressure measurements and yellow pixels indicate sensors with high-pressure measurements.}
    \label{fig:nineImg}
\end{figure*}

The network was implemented and trained on TensorFlow \cite{abadi2016tensorflow} using an NVIDIA Titan Xp. The trainable parameters $\theta$ were initialized randomly and a hessian-based optimization algorithm, L-BFGS \cite{byrd1995limited}, was utilized for minimization of cost function. The training was stopped after $14,500$ epochs since no further improvement was observed beyond this point. The limited-memory BFGS algorithm outperformed non-hessian-based optimization methods such as Adam \cite{kingma2014adam}, in terms of training speed and accuracy (by nearly $3\%$ higher performance) for both tasks (BMI and subject identity). The architecture and parameters of the network were tuned empirically with the goal of maximizing both accuracies. We did not use any regularization other than weight decay with a coefficient of $0.0001$, which was set empirically throughout our experiments to maximize performance.

\subsection{Evaluation and Comparison} \label{sec:sub_sec_eval}
To rigorously test our method, we experimented with both discriminative and generative approaches for BMI prediction. First, we compared the performance of our proposed model to $4$ other generative techniques, namely linear regressor, SVM, Bagged Tree Ensemble (BTE) regression, and Gaussian Process Regression (GPR). Additionally, we compared our proposed solution to $4$ models trained as classifiers, including NB, BTE, $k$NN (with different distance measures), and SVM with different kernels.

In order to utilize discriminative techniques for BMI estimation, the target data are required to be in distinct classes. To achieve this structure with our continuous BMI values, we utilized $k$-means clustering to create a number of classes. $k$-means is an unsupervised algorithm where $k$ is the number of desired clusters, and data points are assigned to classes using a distance function. In this study, we set $k = 5$ based on the distribution of calculated BMI values. Successive to clustering of the data, the $4$ discriminative classifiers mentioned above (SVM, $k$NN, NB, and BTE) were used for evaluation and comparison to our proposed method for distinguishing the BMI classes. An SVM classifies by generating a hyperplane between two classes of data and optimizing the parameters such that the margins around the hyperplane are maximized. $k$NN classifies an input sample by searching through the $k$ nearest training data points. We experimented with a number of distance functions to determine the optimal distance function. Next, we employed NB, a probabilistic method based on the Bayes theorem, and bagged tree ensemble, an ensemble decision tree, as well for comparison. As discussed earlier, we also compared our method to other generative models which can estimate BMI values with higher resolution. For this purpose, a linear model, SVM with a Gaussian kernel, GPR, and BTE were employed. The parameters for both the discriminative and generative benchmark models were tuned empirically through experimentation, with the goal of maximizing accuracy.

\begin{figure}[!t]
    \centering
    \includegraphics[width=0.95\linewidth, scale=0.6]{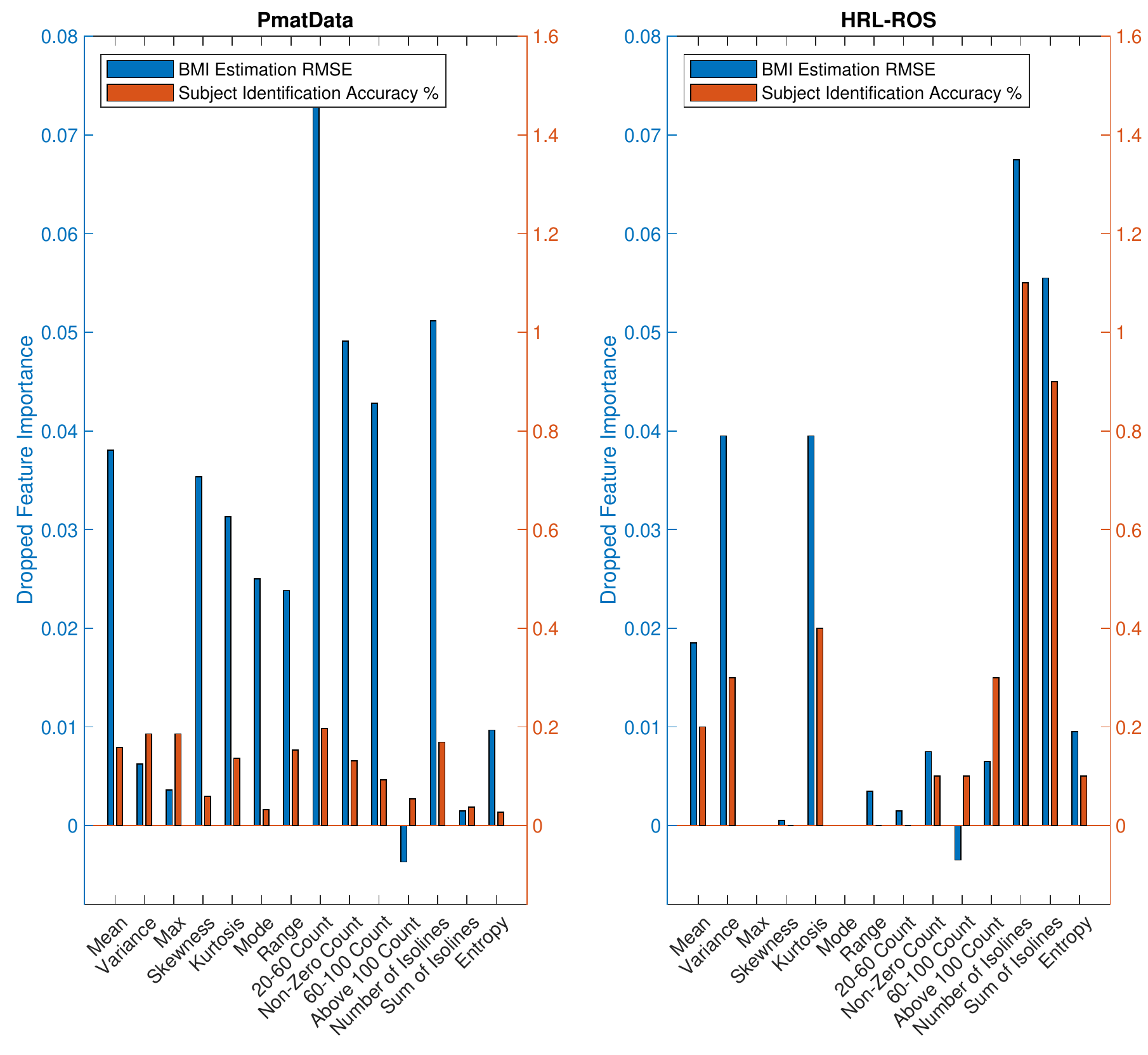}
    \caption{The feature importance, calculated as the difference observed when each feature is dropped, is demonstrated for both datasets.}
    \label{fig:feature_analysis}
\end{figure}

\section{Results and Discussion} \label{sec:sec5}
The pre-processing step produced sharp, contrasting pressure distribution images. Pressure maps for $3$ different subjects in $17$ different postures from the PmatData dataset are presented in Figure \ref{fig:nineImg}. In this figure, participants with higher BMI values appear to have an overall larger area of sensor activation. For example, the subject in the third row appears to have a larger figure with higher pressure points compared to the other two subjects, whereas the subject in the first row seems to have a more petite figure with less pressure applied to the sensors. The same effect was also observed in the HRL-ROS dataset.

Throughout the next sections, first, we present a brief analysis of the extracted features, then we describe the results obtained for each of the two tasks: BMI estimation and user identification using both datasets.

\begin{table}[t]
\centering
\caption{The performance of BMI prediction for PmatData is presented. Values are presented as mean $\pm$ std.}
\label{table_regression}
\begin{center}
\footnotesize
\centering
\scalebox{1} {
\begin{tabularx}{0.9\columnwidth}{lYY}
\hline
\textbf{Model} & \textbf{$R^2$ (\%)}  & \textbf{$RMSE$} \\
\hline
\hline
Linear Regression Model              & 49.4$\pm$3.63       &  3.00$\pm$0.11 \\
BTE                       & 92.1$\pm$0.61       &  1.19$\pm$0.04 \\
SVM (Guassian kernel)     & 95.1$\pm$0.84       &  0.93$\pm$0.08 \\
GPR                       & 96.3$\pm$0.38       &  0.81$\pm$0.04 \\
\textbf{Ours (DNN)}       & \textbf{98.3$\pm$0.37}       &  \textbf{0.55$\pm$0.06} \\
\hline
\end{tabularx}
}
\vspace{-5mm}
\end{center}
\end{table}

\begin{table}[t]
\centering
\caption{The performance of BMI prediction for HRL-ROS is presented. Values are presented as mean $\pm$ std.}
\label{table_regression_d2}
\begin{center}
\footnotesize
\centering
\scalebox{1} {
\begin{tabularx}{0.9\columnwidth}{lYY}
\hline
\textbf{Model} & \textbf{$R^2$ (\%)}  & \textbf{$RMSE$} \\
\hline
\hline
Linear Regression Model              & 80.8$\pm$0.52       &  1.54$\pm$0.03 \\
BTE                       & 89.8$\pm$0.58       &  1.13$\pm$0.03 \\
SVM (Guassian kernel)     & 90.1$\pm$0.91       &  1.08$\pm$0.06 \\
GPR                       & 90.4$\pm$0.54       &  1.08$\pm$0.05 \\
\textbf{Ours (DNN)}       & \textbf{94.9$\pm$0.43}       &  \textbf{0.80$\pm$0.03} \\
\hline
\end{tabularx}
}
\vspace{-5mm}
\end{center}
\end{table}

\begin{table}[t]
\caption{Determining the most effective classes created by $k$-means clustering using the accuracy of an SVM classifier.}
\vspace{-4mm}
\label{table_cluster_results}
\begin{center}
\scalebox{1} {
\begin{tabularx}{0.99\columnwidth}{lYYY}
\hline
\textbf{Input} & $k$ & \textbf{Training} & \textbf{Validation} \\
\hline
\hline
BMI & 5 & 94.1 & 92.6 \\
Age vs. BMI & 4 & 95.4 & 93.1 \\
Weight vs. Height & 5 & 97.1 & 95.6 \\
\hline
\end{tabularx}
}
\vspace{-3mm}
\end{center}
\end{table}

\begin{table}
\caption{The accuracy of BMI classification for PmatData is presented. Values are presented as mean $\pm$ std (in \%).}
\label{table_classification}
\begin{center}
\scalebox{1} {
\begin{tabularx}{0.95\columnwidth}{lY}
\hline
\multirow{1}{*}{\textbf{Classifier}} & \textbf{Validation Accuracy} \\
\hline \hline
NB                           & 52.1 $\pm$ 2.21           \\
$k$NN (Euclidean)            & 96.6 $\pm$ 0.37           \\
$k$NN (Cosine)               & 96.8 $\pm$ 0.34           \\
$k$NN (Cubic)                & 96.1 $\pm$ 0.53           \\
SVM (Linear)                 & 61.8 $\pm$ 1.53           \\
SVM (Gaussian)               & 96.1 $\pm$ 0.21           \\
SVM (Cubic)                  & 83.2 $\pm$ 0.74           \\
BTE                          & 97.4 $\pm$ 0.31           \\
\textbf{Ours (DNN)}          & \textbf{98.8 $\pm$ 0.12}  \\
\hline
\end{tabularx}
}
\end{center}
\end{table}

\begin{table}
\caption{The accuracy of BMI classification for HRL-ROS is presented. Values are presented as mean $\pm$ std (in \%).}
\label{table_classification_d2}
\begin{center}
\scalebox{1} {
\begin{tabularx}{0.95\columnwidth}{lY}
\hline
\multirow{1}{*}{\textbf{Classifier}} & \textbf{Validation Accuracy} \\
\hline \hline
NB                           & 78.7 $\pm$ 2.84           \\
$k$NN (Euclidean)            & 96.4 $\pm$ 0.37           \\
$k$NN (Cosine)               & 92.2 $\pm$ 0.31           \\
$k$NN (Cubic)                & 94.1 $\pm$ 0.53           \\
SVM (Linear)                 & 86.9 $\pm$ 1.22           \\
SVM (Gaussian)               & 96.5 $\pm$ 0.20           \\
SVM (Cubic)                  & 96.4 $\pm$ 0.17           \\
BTE                          & 96.4 $\pm$ 0.32           \\
\textbf{Ours (DNN)}          & \textbf{97.4 $\pm$ 0.18}  \\
\hline
\end{tabularx}
}
\end{center}
\end{table}

\begin{table*}[t]
\centering
\caption{User identification evaluation for each subject on PmatData (in \%).}
\label{table_eval}
\begin{center}
\centering
\begin{tabularx}{1\textwidth}{lYYYYYYY}
\hline 
 Subject &  1 &  2 &  3 &  4 &  5 &  6 &  7 \\
\hline 
\hline
 Precision &  97.7$\pm$1.0   &   98.2$\pm$1.6   &   98.0$\pm$1.2   &   98.0$\pm$1.4   &   98.5$\pm$1.3   &   98.6$\pm$1.6   &   99.1$\pm$0.6  \\
 Recall    &  97.7$\pm$1.3   &   97.7$\pm$1.7   &   98.5$\pm$0.7   &   98.2$\pm$1.0   &   98.6$\pm$1.2   &   98.5$\pm$1.2   &   98.8$\pm$0.3  \\
 F1-Score  &  97.7$\pm$0.8   &   97.9$\pm$1.2   &   98.3$\pm$0.6   &   98.1$\pm$0.8   &   98.5$\pm$0.8   &   98.5$\pm$1.0   &   99.0$\pm$0.3  \\ 
\hline 
 Subject &  8 &  9 &  10 &  11 &  12 &  13 & Average \\
\hline 
\hline
 Precision &  99.0$\pm$1.0   &   98.5$\pm$1.2   &   97.3$\pm$1.4   &   98.2$\pm$1.0   &   98.9$\pm$0.6   &   97.6$\pm$1.3   &   98.3$\pm$0.2 \\
 Recall    &  99.0$\pm$0.7   &   98.8$\pm$1.1   &   97.8$\pm$1.5   &   98.1$\pm$1.4   &   98.6$\pm$1.6   &   97.4$\pm$1.6   &   98.3$\pm$0.2 \\
 F1-Score  &  99.0$\pm$0.6   &   98.7$\pm$0.7   &   97.6$\pm$0.8   &   98.1$\pm$0.5   &   98.7$\pm$0.8   &   97.5$\pm$0.7   &   98.3$\pm$0.2 \\ 
\hline 
\end{tabularx}
\end{center}
\end{table*}

\begin{table*}[t]
\centering
\caption{User identification evaluation for each subject on HRL-ROS (in \%).}
\label{table_eval_d2}
\begin{center}
\centering
\begin{tabularx}{1\textwidth}{lYYYYYYYYY}
\hline 
 Subject &  1 &  2 &  3 &  4 &  5 &  6 &  7 &  8 &  9 \\
\hline 
\hline
 Precision   &   97.9$\pm$1.1   &   96.5$\pm$0.7   &   89.0$\pm$1.8   &   88.5$\pm$3.4   &   98.0$\pm$0.7   &   97.0$\pm$1.5   &   99.5$\pm$0.4   &   85.8$\pm$1.6   &   99.0$\pm$0.6 \\
 Recall      &   97.6$\pm$0.5   &   98.2$\pm$0.8   &   92.7$\pm$2.2   &   75.4$\pm$2.3   &   99.6$\pm$0.3   &   93.4$\pm$2.0   &   98.7$\pm$0.9   &   92.6$\pm$1.7   &   97.7$\pm$0.9 \\  F1-Score    &   97.8$\pm$0.7   &   97.3$\pm$0.5   &   90.8$\pm$1.1   &   81.4$\pm$2.4   &   98.8$\pm$0.4   &   95.2$\pm$1.3   &   99.1$\pm$0.6   &   89.0$\pm$1.3   &   98.3$\pm$0.6 \\ 
\hline 
 Subject &  10 &  11 &  12 &  13  &  14  &  15  &  16  &  17   &   Average\\
\hline 
\hline
 Precision   &   99.3$\pm$0.5   &   95.3$\pm$1.8   &   99.6$\pm$0.4   &   97.3$\pm$0.5   &   99.7$\pm$0.3   &   99.0$\pm$0.3   &   99.1$\pm$0.9   &   85.5$\pm$4.0   &   95.7$\pm$0.4 \\
 Recall      &   99.7$\pm$0.3   &   98.0$\pm$1.1   &   99.4$\pm$0.7   &   97.6$\pm$1.6   &   99.4$\pm$0.5   &   99.1$\pm$0.4   &   97.6$\pm$0.7   &   79.2$\pm$3.1   &   95.0$\pm$0.4 \\
 F1-Score    &   99.5$\pm$0.3   &   96.6$\pm$1.0   &   99.5$\pm$0.3   &   97.4$\pm$0.8   &   99.5$\pm$0.3   &   99.1$\pm$0.3   &   98.3$\pm$0.5   &   82.1$\pm$2.2   &   95.3$\pm$0.4 \\ 
\hline 
\end{tabularx}
\end{center}
\end{table*}

\subsection{Feature Analysis}
In order to investigate the extracted features, we used a \textit{drop-column feature importance} strategy to measure their contributions to the two tasks. In this method, we evaluate the importance of each feature by comparing a model using all of the input features against the same model with the feature in question removed. Although this process is quite time-consuming, reliable results are generally achieved. The accuracy differences obtained between the full feature set and the scenarios where individual features are dropped, are demonstrated in Figure \ref{fig:feature_analysis}, where on most cases, removing any feature from the PmatData dataset yields worst results for both tasks, meaning that all of the features are contributing to both tasks. It should be noted that for the HRL-ROS dataset, as the published data were pre-normalized between $0-1024$, \textit{max} and \textit{range} features were not utilized in our model. Based on the results, the rest of the $12$ remaining features have all important contributions towards the performance. It can also be noted that on both datasets, while some features are more important than others, in the absence of a single feature, other features seem to compensate to a good degree. For each dataset, only one feature has a negative impact on BMI estimation: the number of pixels above $100$ in PmatData, and the number of pixels between $60$ and $100$ in HRL-ROS. Since these feature do contribute to subject identification and are also different for the two datasets, we decided to keep them as one of our utilized features. Lastly, since the final BMI estimation and identity recognition results (shown in the following sections) are very accurate and consistent, further features were not explored.


\subsection{Task 1: BMI Estimation}
\subsubsection{Generative Models}
A $10$-fold cross-validation scheme was used to evaluate the different models, including our proposed deep multitask neural network. In order to evaluate generative models, the $R^2$ and root mean square error (RMSE) metrics are used. $R^2$ is a useful metric that demonstrates both fitted bias and variance of the regression. We also considered measuring the mean percentage error (MPE) and mean absolute percentage error (MAPE). However, differences in the generative models were not captured clearly using such metrics. Tables \ref{table_regression} and \ref{table_regression_d2} present the results of our method compared to the benchmark generative models on PmatData and HRL-ROS respectively. The results show that almost all of the methods achieve high accuracies, indicating that the models generalize well for the set of extracted features. On the PmatData dataset, it is observed that our proposed solution achieves an $R^2$ score of $98.3$ and RMSE of $0.55$, considerably outperforming other techniques. Similarly, on the HRL-ROS dataset, we observe a significant improvement of performance with an $R^2$ score of $94.9$ and an RMSE of $0.8$. Furthermore, based on the results, we conclude that linear regression models are not appropriate for BMI estimation.

Our method shows higher performance on PmatData compared to HRL-ROS dataset, as the PmatData dataset has a smaller variety of subjects and postures. However, on both datasets, our model has an average RMSE of less than one, which can be considered a good performance as BMI categories studied in the literature have a range of $5$ on average.

\subsubsection{Discriminative Models}
For further evaluation of our proposed regression model for BMI estimation, discriminative algorithms (classifiers) were also utilized. As discussed earlier, the BMI values were clustered using $k$-means into a number of distinct classes. Clustering was performed on BMI alone, as well as BMI vs. age, and weight vs. height, with $k = 1, ..., 6$. We identified the most effective number of classes by observing the accuracy of an SVM classifier on each cluster. The results of this experiment on PmatData is shown in Table \ref{table_cluster_results}. On both datasets, the best results were obtained for $k = 5$ using weight vs. height, which is interestingly also equal to the $5$ categories used in the literature for categorizing BMI levels \cite{cnattingius1998prepregnancy}.

Successive to clustering, the mentioned discriminative models were trained using the extracted features as inputs and the $5$ classes (clusters) as outputs. The parameters of the classifiers were tuned empirically with the goal of maximizing classification accuracy. For the $k$NN classifier, the best accuracy was achieved with $k = 10$. Additionally, a number of different distance metrics were explored for the $k$NN. We experimented with Cosine distance {($d(a,b) = \frac{a.b}{\sqrt{a.a}\sqrt{b.b}}$)}, Minkowski (Cubic) {($d(a,b) = \left(\sum_{i=1}^n |a_i-b_i|^3\right)^{1/3}$}, and Euclidean distances. For SVM, different kernels including linear, Gaussian, and Cubic were utilized. Lastly, a BTE and a NB classifier were trained as discussed in Section \ref{sec:sub_sec_eval}.

\begin{table*}[!tb]
\caption{Subject identification performance comparison for PmatData in $3$ posture categories.}
\label{table_comparison}
\begin{center}
\begin{tabularx}{\textwidth}{l|YYY|YYY|YYY}
\hline
\multirow{2}{*}{\textbf{Method}}     &               &   Supine       &                &                &    Left        &                &                &   Right        &                  \\
\cline{2-10}
                                     &  Accuracy     &  Precision     &  Recall        &  Accuracy      &  Precision     &  Recall        &  Accuracy      &  Precision     &  Recall          \\
\hline
\hline
NB &  24.2$\pm$2.2  &  29.8$\pm$4.1  &  28.1$\pm$1.4  &  22.3$\pm$2.2  &  18.1$\pm$3.1  &  25.6$\pm$1.3  &  20.3$\pm$2.7  &  29.9$\pm$4.4  &  24.0$\pm$1.3    \\
$k$NN (Euclidean)                   &  96.2$\pm$0.3  &  96.9$\pm$0.4  &  96.1$\pm$0.3  &  96.2$\pm$0.8  &  96.0$\pm$0.9  &  96.1$\pm$0.9  &  96.6$\pm$0.4  &  96.3$\pm$0.5  &  96.5$\pm$0.3    \\
$k$NN (Cosine)                      &  96.2$\pm$0.4  &  96.9$\pm$0.4  &  96.1$\pm$0.3  &  96.3$\pm$0.7  &  96.1$\pm$0.8  &  96.2$\pm$0.8  &  96.6$\pm$0.4  &  96.3$\pm$0.5  &  96.5$\pm$0.3    \\
$k$NN (Cubic)                       &  96.1$\pm$0.3  &  96.8$\pm$0.4  &  96.0$\pm$0.3  &  96.2$\pm$0.7  &  96.1$\pm$1.0  &  96.1$\pm$0.8  &  96.6$\pm$0.4  &  96.4$\pm$0.4  &  96.5$\pm$0.3    \\
SVM (Linear)                        &  70.2$\pm$1.3  &  71.4$\pm$1.3  &  67.2$\pm$1.2  &  62.8$\pm$1.7  &  63.6$\pm$1.9  &  59.1$\pm$2.0  &  60.2$\pm$1.7  &  57.2$\pm$2.0  &  58.2$\pm$1.6    \\
SVM (Gaussian)                      &  96.9$\pm$0.3  &  96.6$\pm$0.4  &  96.9$\pm$0.3  &  96.8$\pm$0.6  &  96.8$\pm$0.5  &  96.6$\pm$0.6  &  96.9$\pm$0.7  &  96.7$\pm$0.7  &  96.7$\pm$0.7    \\
SVM (Cubic)                         &  95.9$\pm$0.5  &  95.8$\pm$0.5  &  95.5$\pm$0.5  &  94.6$\pm$0.7  &  95.5$\pm$0.7  &  94.6$\pm$0.8  &  95.1$\pm$1.1  &  95.5$\pm$1.0  &  94.6$\pm$1.0    \\
BTE                                 &  96.7$\pm$0.3  &  96.4$\pm$0.4  &  96.5$\pm$0.4  &  96.8$\pm$0.6  &  96.7$\pm$0.6  &  96.6$\pm$0.8  &  96.5$\pm$0.6  &  96.3$\pm$0.7  &  96.2$\pm$0.7    \\
\cite{pouyan2017pressure}           &  85.5          &  85.9          &  85.8          &  80.4          &  76.0          &  80.0          &  82.3          &  83.3          &  81.9            \\  
\textbf{Ours (DNN)}                 &  \textbf{98.4$\pm$0.3}  &  \textbf{98.2$\pm$0.3}  &  \textbf{98.3$\pm$0.3}  &  \textbf{98.4$\pm$0.5}  &  \textbf{98.4$\pm$0.6}  &  \textbf{98.3$\pm$0.4}  &  \textbf{98.4$\pm$0.4}  &  \textbf{98.3$\pm$0.4}  &  \textbf{98.2$\pm$0.4}    \\
\hline
\end{tabularx}
\vspace{-3mm}
\end{center}
\end{table*}

As presented in Table \ref{table_classification}, for PmatData, the BTE showed the best results among the benchmarks. In particular, the liner kernel showed poor accuracy, indicating that similar to the generative linear regressor, the classes are not linearly separable. Similarly, when evaluating our method on HRL-ROS dataset, BTE showed the highest accuracy, closely followed by SVM (cubic) and $k$NN.

To convert the BMI outputs of our proposed generative model to $5$ discrete classes, we used a logistic regression classifier on the extracted feature representations from the fifth hidden dense layer to make the prediction. As presented in Table \ref{table_classification}, our proposed solution provided the best validation accuracy of $98.3\%$. Furthermore, as seen in Table \ref{table_classification_d2}, our proposed DNN also outperforms the benchmarks by achieving a validation accuracy of $97.4\%$. Besides achieving the top performance, another advantage of our model over discriminative methods is predicting continuous values which can be more useful for monitoring health quality over time.

\subsection{Task 2: Subject Identification}
\begin{figure}[t]
    \centering
    \includegraphics[width=0.95\linewidth, scale=0.6]{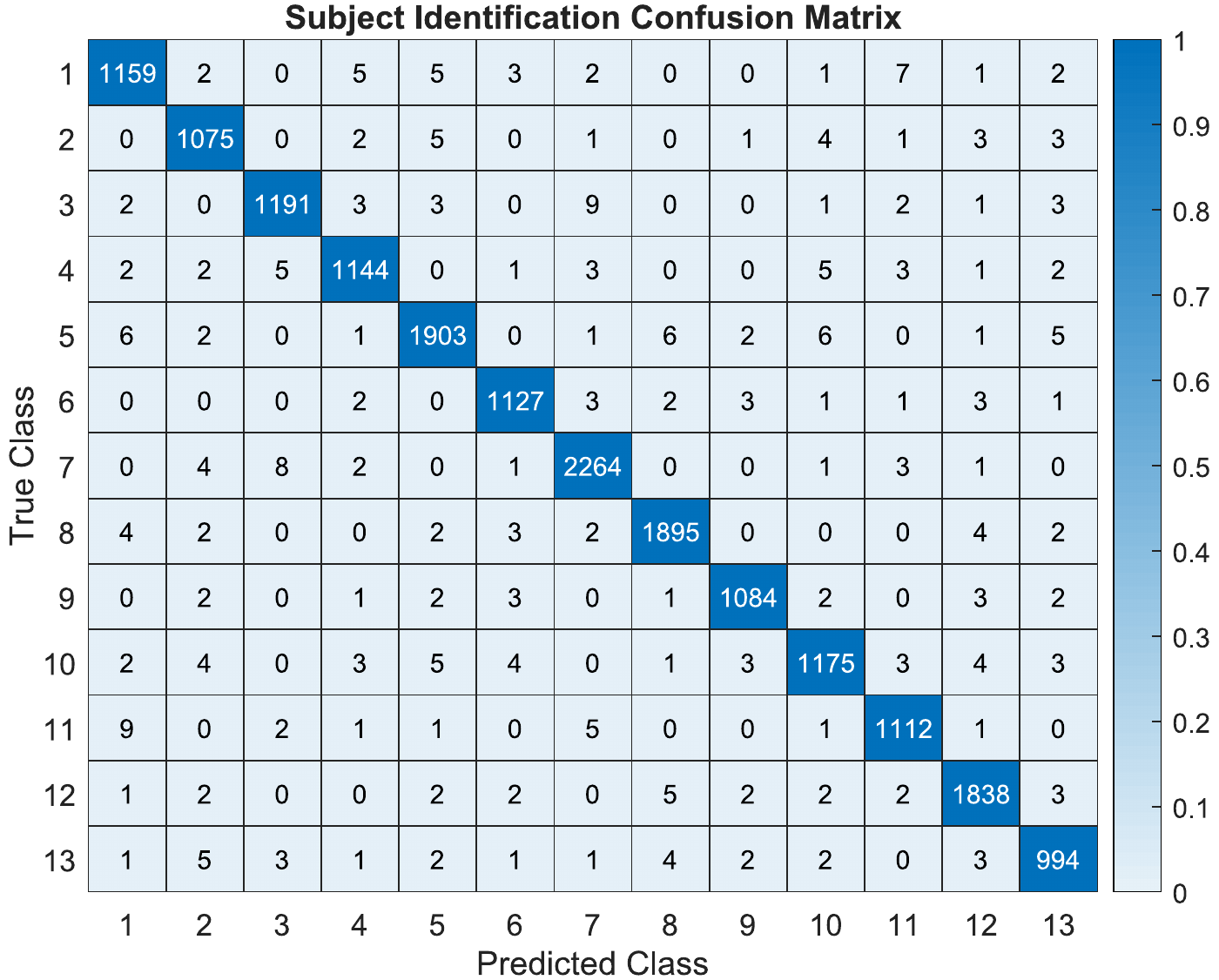}
    \caption{Confusion matrix for user identification on PmatData.}
    \label{fig:conf}
\end{figure}

\begin{figure}[t]
    \centering
    \includegraphics[width=0.95\linewidth, scale=0.6]{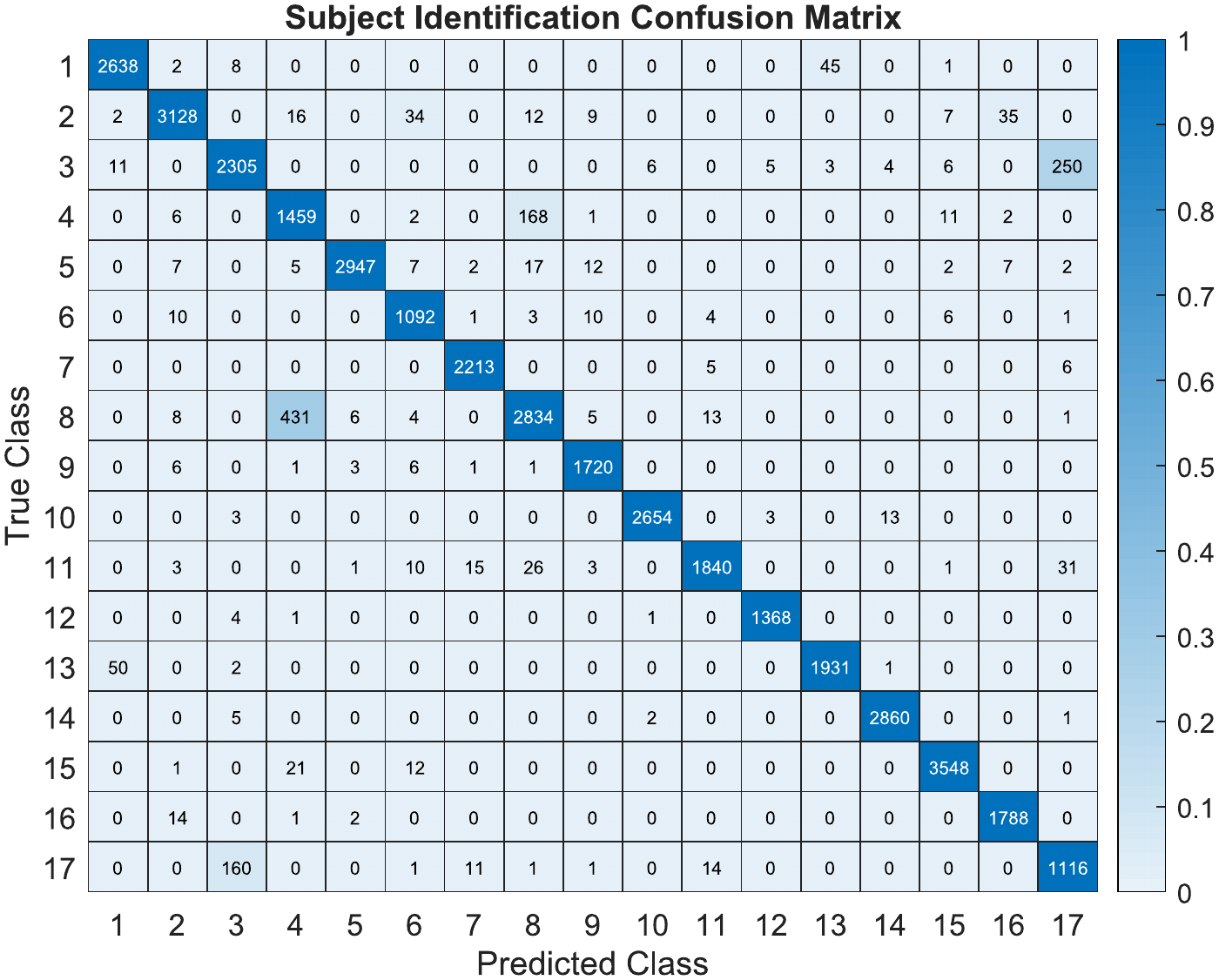}
    \caption{Confusion matrix for user identification on HRL-ROS.}
    \label{fig:conf_d2}
\end{figure}

Our proposed DNN was able to identify subjects with a $10$-fold validation accuracy of $98.4\%$ on the PmatData dataset and $95.7\%$ on the HRL-ROS dataset. Our network also achieved a close performance on the training set, showing a small difference (under $2\%$) between validation and training, which can be interpreted as almost no over-fitting. Figures \ref{fig:conf} and \ref{fig:conf_d2} show the resulting identification confusion matrices for the two datasets based on our proposed solution. To further demonstrate our model's performance, the precision, recall, and F1-Score values for each subject were also calculated for both datasets and summarized in Tables \ref{table_eval} and \ref{table_eval_d2}, where a consistent performance for all the subjects and datasets is observed.

\begin{table}[t]
\caption{Subject identification performance comparison for HRL-ROS.}
\label{table_comparison_d2}
\begin{center}
\begin{tabularx}{\columnwidth}{l|YYY}
\hline
\textbf{Method}                    &    Accuracy     &  Precision     &  Recall          \\
\hline
\hline
NB  &  68.8$\pm$0.7  &  65.7$\pm$2.3  &  67.4$\pm$0.7    \\ 
$k$NN (Euclidean)                   &  91.6$\pm$0.5  &  92.4$\pm$0.5  &  90.4$\pm$0.6    \\    
$k$NN (Cosine)                      &  91.5$\pm$0.5  &  92.3$\pm$0.4  &  90.4$\pm$0.4    \\   
$k$NN (Cubic)                       &  91.0$\pm$0.5  &  91.8$\pm$0.5  &  89.8$\pm$0.5    \\         
SVM (Linear)                        &  84.8$\pm$0.6  &  85.0$\pm$0.5  &  82.9$\pm$0.5    \\        
SVM (Gaussian)                      &  93.7$\pm$0.5  &  94.3$\pm$0.5  &  92.7$\pm$0.4    \\      
SVM (Cubic)                         &  92.6$\pm$0.4  &  93.2$\pm$0.4  &  91.3$\pm$0.3    \\    
BTE                                 &  93.4$\pm$0.4  &  93.2$\pm$0.5  &  92.4$\pm$0.4    \\    
\textbf{Ours (DNN)}                 &  \textbf{95.8$\pm$0.3}  &  \textbf{95.7$\pm$0.4}  &  \textbf{95.0$\pm$0.4}    \\
\hline
\end{tabularx}
\vspace{-3mm}
\end{center}
\end{table}

In Table \ref{table_comparison}, we compare our results on the PmatData to the same machine learning benchmarks used for BMI estimation, along with those reported in \cite{pouyan2017pressure}, where subject identification was performed on the same dataset in $3$ posture categories (left, right, and supine). \cite{pouyan2017pressure} trained separate models for each posture group, while our method utilized only one model to identify users in all postures over a total of $10$ postures. This, we believe is a significant advantage of our proposed method as we were able to classify people in a wider range of postures (a more realistic representation of natural sleeping postures), with higher accuracy and less complexity. Overall, our model performed better compared to the previous study for each posture and we achieved a relatively more consistent classification result among the $3$ posture categories. Specifically, we achieved an average accuracy of $98.4\%$ across the $3$ postures, outperforming \cite{pouyan2017pressure}, which had an average accuracy of $82.7\%$. Furthermore, since the benchmark models also achieve better performance compared to \cite{pouyan2017pressure}, we can conclude that the set of extracted features used in our method were very suitable for the experiments. For the HRL-ROS dataset, since there are no prior works on subject identification, we also provide the performance of other methods in Table \ref{table_comparison_d2}. Similar to the discriminative models for BMI estimation, the linear models and NB are not able to compete with other methods, where SVM with Gaussian kernel and BTE show the best accuracy after our proposed method. Since this dataset contains a large variation of lying postures, categorizing them into the $3$ sleeping categories was not possible for this dataset.

\section{Conclusion} \label{sec:sec6}
In this paper, we proposed a method for robust pervasive estimation of BMI and identification of subjects simultaneously in smart-beds with integrated textile-based pressure sensors. For this purpose, we first performed noise reduction in both time and spatial domains. Next, we extracted $14$ different features from each frame of the pressure matrices. The ground-truth BMI values were calculated using the weight and height values available in the dataset. We then developed a deep multitask neural network capable of estimating the BMI values and identifying subjects with high accuracy. For evaluation purposes, a number of other methods were also used, including different discriminative and generative models. Our experiments on two public datasets, PmatData and HRL-ROS, showed that the proposed DNN model outperformed the other methods in a $10$-fold cross-validation scheme and performed with high accuracy of $98.8\%$ and $97.4\%$ for estimating BMI respectively. We were able to estimate BMI reliably in a variety of postures, which indicates the potential of the network to reliably predict a user's BMI in any position. Furthermore, we performed identity recognition for the subjects. Our multi-tasking approach outperformed the state of the art method that had utilized deep learning \cite{pouyan2017pressure}. Additionally, our approach improved upon the generalizability of previous subject identification efforts. Our proposed model was able to accurately identify subjects in all the available positions. 
Our work has various applications for personalized health in smart homes and potentially smart hospitals, where automated detection and recording of BMI can save time and effort, and will most likely result in more accurate values compared to self-reporting due to the minimization of human errors. Moreover, users seeking weight gain or weight loss can keep track of small changes in their BMI, which may have considerable implications for both health and motivation. Finally, our approach can be utilized for user identification systems, resulting in personalized ambient conditions, such as lighting and sound, and bed position to the user's specific preferences.

\section*{Conflict of Interest}
The authors declare that they have no conflict of interest.

\section*{Acknowledgements}
This work was funded by the Natural Sciences and Engineering Research Council of Canada (NSERC), Discovery Grant. The Titan XP GPU used for this research was donated by the NVIDIA Corporation.

\bibliographystyle{IEEEtran}
\bibliography{IEEEabrv,main}

\end{document}